\documentclass[11pt]{article}
\usepackage[utf8]{inputenc}
\usepackage{cite}
\usepackage{booktabs}
\usepackage{algorithm}
\usepackage{algorithmic}
\usepackage{xspace}
\usepackage{pgf}
\usepackage{multirow}
\usepackage{xcolor}
\usepackage{stfloats}

\usepackage{authblk}
\usepackage{rotating}
\usepackage{fullpage}
\usepackage{amsmath, amssymb}
\usepackage{hyperref}
\hypersetup{
	hidelinks,
	colorlinks=true,
	citecolor=[rgb]{0.121 0.47 0.705},
	linkcolor=[rgb]{0.121 0.47 0.705},
	urlcolor=[rgb]{0.121 0.47 0.705}
}

\usepackage{todonotes}

\newcommand{\qgis}{\textsc{QGIS}\xspace}
\newcommand{\falp}{\textsc{FALP}\xspace}
\newcommand{\chain}{\textsc{CHAIN}\xspace}
\newcommand{\popmusic}{\textsc{POPMUSIC}\xspace}
\newcommand{\simple}{\textsc{GREEDY}\xspace}
\newcommand{\indset}{\textsc{MIS}\xspace}
\newcommand{\maxhs}{\textsc{MHS}\xspace}

\newcommand{\kamis}{\textsc{KAMIS}\xspace}

\newtheorem{problem}{Problem}

\title{Exploring Semi-Automatic Map Labeling}
\author[1]{Fabian Klute} %
\author[1]{Guangping Li} %
\author[1]{Raphael L\"offler} %
\author[1]{Martin N\"ollenburg} %
\author[2]{Manuela~Schmidt} %

\affil[1]{Algorithms and Complexity Group, TU Wien, Vienna, Austria}
\affil[2]{Research Group Cartography, TU Wien, Vienna, Austria}

\date{}

\begin{document}

	\maketitle

	\maketitle

	\begin{abstract}
		Label placement in maps is a very challenging task that is critical for the overall map quality.
		Most previous work focused on designing and implementing fully automatic solutions, but the resulting visual and aesthetic quality has not reached the same level of sophistication that skilled human cartographers achieve. 
		We investigate a different strategy that combines the strengths of humans and algorithms.
		In our proposed method, first an initial labeling is computed that has many well-placed labels but is not claiming to be perfect. Instead it serves as a starting point for an expert user who can then interactively and locally modify the labeling where necessary. %
		In an iterative human-in-the-loop process alternating between user modifications and local algorithmic updates and refinements the labeling can be tuned to the user's needs.
		
		We demonstrate our approach by performing different possible modification steps in a sample workflow with a prototypical interactive labeling editor. 
		Further, we report computational performance results from a simulation experiment in QGIS, which investigates the differences between exact and heuristic algorithms for semi-automatic map labeling. 
		To that end, we compare several alternatives for recomputing the labeling after local modifications and updates, as a major ingredient for an interactive labeling editor.
	\end{abstract}

	\section{Introduction}

	Label placement is an important step in map production, both manual and automatic, and it can require up to 50 percent of the total map production time for manually created maps~\cite{yoeli1972}. Imhof's 1975 statement ``Good form and placing of type make the good map. Poor, sloppy, amateurish type placement is irresponsible; it spoils even the best image and impedes reading.''~\cite{imhof1975} has not lost its validity until today. Yet, with more and more automation in cartography and fully digital map production, one can argue that the quality of label placement has not improved or even diminished compared to skilled, but tedious manual label placement~\cite{s-gpusye-12}.
	While practical label placement algorithms are typically very fast and can compute overlap-free positions of thousands of labels within seconds, the resulting maps usually do not meet the highest quality standards but must be carefully post-processed in tedious work by human cartographers.
	There is a lack of algorithmic support of human involvement in an automated labeling workflow. 
	Ideally, a responsive labeling algorithm would be able to react on human interaction and respect any added constraints, e.g., choosing an alternative position of a label or changing its font size, by locally updating the solution while keeping the already placed labels as stable as possible. 
	Such a semi-automatic map labeling tool would allow for a much more comfortable and intelligent human-in-the-loop label placement process in digital map production, where neither human alone can deal with the full data complexity, nor machine alone can deal with the fine-tuning and optimization of mathematically somewhat ill-defined map aesthetics. 
	Instead, the tool should combine the computational power and mathematical rigor of geometric labeling algorithms with the expertise and sense of aesthetics of experienced domain experts in cartography.
	
	In this paper we present a prototype for a such a human-in-the-loop label placement approach that supports first applying selected labeling algorithms for placing an initial set of non-overlapping labels for point features that, in our case, maximizes an objective function counting the (weighted) number of labeled features. 
	To proceed from this initial solution our prototype implements several editing and interaction tools for modifying the labeling according to the needs of an expert user, e.g., changing the visibility or position of a label, or its size, shape, and weight.
	Upon each of those modifications, a second algorithm is re-optimizing and refining the labeling by taking into account both the user input and the existing solution so as to satisfy the new constraints and maximize the stability with respect to the previous solution. 
	That is, a labeling is computed that contains as many as possible of the previously displayed labels and at the same time resolves any new label conflicts resulting from the user input.
	Our prototype is designed to be flexible as to which algorithm to actually use for computing initial solutions and iterative updates. 
	Secondly, we take an algorithm engineering perspective on the map labeling problem and perform an experimental simulation study in the GIS software QGIS. 
	The aim is to investigate differences between and suitability of heuristic and exact algorithms (including those provided by QGIS itself) for the envisioned interactive labeling workflow, which requires frequent recomputation of labelings after local modifications.

		\paragraph{Related Work}
Based on general cartographic guidelines~\cite{imhof1975, w-digtpssm-00}, the first algorithmic solutions to the label placement problem have been studied in the cartographic literature in the 1970s and 1980s~\cite{yoeli1972,hirsch82,z-ipalpp-86}. In the early 1990s the problem has been introduced as a geometric independent set problem to the computational geometry community~\cite{Formann1991,ms-ccclp-91}, where it was recognized as an important application challenge in the computational geometry task force report~\cite{c-accgitfr-96}. 
	It was quickly shown that almost all variants of label placement and label selection problems are NP-hard~\cite{Formann1991,ms-ccclp-91}.
	Therefore, researchers focused on special cases, approximation algorithms and heuristics for label number maximization or label size maximization problems, predominantly for point features, e.g.~\cite{ww-pla-95,wwks-trsglp-01,vanKreveld1998,s-gaclp-01,christensen1995}; for surveys and general introductions see, e.g., \cite{s-gaclp-01,kt-la-13,ws-mlb-09}. 
	More recent works introduced advanced multi-criteria optimization models~\cite{dksw-teqnpm-02,rr-cmmhcqplp-14,hw-bmieipfpl-17} that can express more accurately several established cartographic principles, but still with the aim of a full automation of the map labeling process.
	While progress is made by incorporating more comprehensive  cartographic rules for label placement, none of the above approaches includes decisions made by human experts -- other than setting  preferences, parameters, and priorities in the different scoring functions that control a single optimization run of the respective algorithm. 
	A notable exception is the UserHints framework~\cite{Nascimento2008}, where human interaction was integrated into solving the label number maximi\-za\-tion problem in a fixed-position point labeling setting. In that system, two heuristic methods were implemented as labeling algorithms, and hence the evaluation could not assess the deviation from optimal solutions with respect to the objective function. Moreover, the authors did not consider the stability of the labeling under user interaction.
	Beyond the label placement problem, interactive optimization~\cite{slk-iho-02} and human-guided search~\cite{klmm-hs-09} are of course techniques that are of general interest and more broadly applicable.

	Popular GIS software like \href{https://mapbox.com}{Mapbox}\footnote{see \url{https://mapbox.com}}, 
	\href{https://pro.arcgis.com}{ArcGIS Pro}\footnote{see \url{https://pro.arcgis.com}}, %
	or \href{https://www.qgis.org}{QGIS}\footnote{see \url{https://www.qgis.org}} %
	also provide labeling al\-go\-rithms.
	Mapbox 
	allows customized label modifications with data conditions, but no manual selection or drag-and-drop placement.
	The ArcGIS Pro documentation\footnote{see \url{https://pro.arcgis.com/en/pro-app/help/mapping/text/labeling-basics.htm}}
	states ``Label positions are generated automatically. Labels are not selectable. You cannot edit the display properties of individual labels.'' To allow for manual adjustment, labels can be converted to annotations. If the labels are stored in a database, the annotations can be feature-linked, i.e., the annotations update in case features are added or changed. However, after converting to annotations, all positioning needs to be done manually.
	Other proprietary software developers like \href{https://1spatial.com/}{1Spatial}\footnote{see \url{https://1spatial.com/}} 
	and \href{http://lorienne.com/en/}{Lorienne}\footnote{see \url{http://lorienne.com/en/}}
	advertise features to modify labels in a more advanced manner.
	Though, there seems to be no focus on how to better integrate the automatic labeling process into a more interactive approach,
	especially from an algorithmic perspective.
	Finally, in QGIS 3 some advanced labeling tools were introduced. 
	For example, it is possible to manually drag and reposition labels; other labels will be re-placed accordingly. 
	Labels, which were manually edited, can be highlighted and reversed to their default position.
	While this is a good example that demonstrates the awareness and practical need for semi-automatic labeling solutions, prior to this paper no experimental studies on the performance of different labeling algorithms under interactive editing 
	have been published that evaluate such an approach and guide further development in QGIS and other systems.

		\paragraph{Paper Structure}
	In Section~\ref{sec:model} we introduce our model for semi-automatic map labeling, which combines the classic point-feature label placement with a dynamic update problem. Section~\ref{sec:framework} introduces our prototype tool and describes a sample map labeling workflow using interactive modifications by a cartographic expert.
	Finally, Section~\ref{sec:algexpqgis} describes our simulation experiment in QGIS to analyze the performance of several labeling algorithms.

	\section{Labeling Model}\label{sec:model}
	In this paper we restrict our attention to the \emph{point feature label placement} (PFLP) problem, which is defined as follows. 
	Let $ P $ be a set of $n$ feature points in $ \mathbb{R}^2 $. For each point $p \in P$ we are given a finite set $L_p$ of \emph{label candidates}, where each label candidate $\ell \in L_p$ is represented by the bounding box $R_\ell$ of the feature name placed at a particular position. While in general the label candidates in $L_p$ can be arbitrary label positions, we focus on the standard 4- and 8-position models, where either one of the corners of $R_\ell$ coincides with $p$ (4-position model) or one of the corners or midpoints of the edges of $R_\ell$ coincides with $p$ (8-position model). Let $\mathcal L = \bigcup_{p \in P} L_p$ be the union of all label candidates.

	We say that two label candidates $\ell$ and $\ell'$ are in \emph{conflict}, if the two rectangles $R_\ell$ and $R_{\ell'}$ intersect. Since the names of two conflicting label candidates would overlap, the goal in map labeling is to find a \emph{conflict-free} solution set $\mathcal S \subseteq \mathcal L$ of label candidates. In particular, we require that any two label candidates of the same point $p$ are in conflict so that each point receives at most one label. To optimize the labeling we define a quality function $w \colon \mathcal L \rightarrow \mathbb R^+$ that assigns a weight to each label candidate. Then in its basic form, which we implemented for our prototype tool and is also used equivalently in QGIS through a cost model for non-labeled features, the PFLP optimization problem is defined as follows.
	\begin{problem}[PFLP]
		Given a set $P$ of $n$ points in $\mathbb{R}^2$ with a set of label candidates $\mathcal L$ and a weight function $w$, find a conflict-free set $S \subseteq \mathcal L$ of label candidates for $P$ such that the weight $W(\mathcal S) = \sum_{\ell \in \mathcal S} w(\ell)$ is maximized.
	\end{problem}
	The simplest weight function is $w \equiv 1$, which just counts the number of selected labels. But more advanced weight functions, defined for single labels, pairs of labels, or even larger subsets, in order to model various cartographic principles are possible~\cite{rr-cmmhcqplp-14,hw-bmieipfpl-17,dksw-teqnpm-02}.
	We want to explore user modifications in our semi-automatic labeling process, which, for example, change the set $\mathcal L$ of label candidates to a set $\mathcal L'$ or the weight function $w$ to a function $w'$. Therefore we define the following PFLP update problem.
	\begin{problem}[PFLP-Update]
		\label{prob:PFLP-Update}
		Given a set $P$ of $n$ points in $\mathbb{R}^2$ with a set of label candidates $\mathcal L'$, a weight function $w'$, and a previous labeling $\mathcal S$, compute a conflict-free solution $\mathcal S' \subseteq \mathcal L'$ that maximizes the number $|\mathcal S \cap \mathcal S'|$ of \emph{stable} labels as well as the weight $W(\mathcal S') = \sum_{\ell \in \mathcal S'} w'(\ell)$.
	\end{problem}
	We note that there may be a trade-off between the stability of the new solution $\mathcal S'$ and its weight $W(\mathcal S')$ that can be adjusted by the user.
	For solving the two labeling problems algorithmically, we model the conflicts and the label candidates as a weighted \emph{conflict graph} $G=(V,E)$, where the vertex set $V=\mathcal L$ consists of all label candidates and the edge set $E$ consists of all pairs of conflicting label candidates. Then, in graph-theoretic terms, an optimal labeling corresponds to a \emph{maximum weight independent set} in $G$, i.e., a subset $V' \subseteq V$ of vertices such that no two vertices $u,v \in V'$ are adjacent and the weight $\sum_{v \in V'} w(v)$ is maximum. The problem of computing maximum independent sets in graphs is a classic NP-hard problem~\cite{Garey79}, even in its unweighted form.
	\subsection{Data}\label{sec:data}
	For our computational experiments and the evaluation of the prototype we extracted points-of-interest data from the \href{https://openstreetmap.org}{OpenStreetMap}\footnote{\url{https://openstreetmap.org}} (OSM)  project,  
	filtered it for certain categories or properties, and then stored the name and location of the remaining points as ESRI shapefiles %
	or in a simple JSON file format to be read by our tool. 
	Using data from the OSM project guarantees that the feature distribution is realistic, even if the particular data sets are simplified and not cartographically sound use cases.

	We compiled five different datasets, all with unit weights, whose properties are summarized in Tab.~\ref{tab:data}. The first one consists of all mountain peaks above 2,499 meters in the mountain range “Hohe Tauern” in Austria.
	It consists of  1,278 homogeneously distributed, natural features and is on average less dense than the other four data sets.
	We compiled this dataset to use it in the sample workflow with a zoom level set to 12, see Section~\ref{sec:sampleworkflow}. 
	The number of conflicts and conflicts per feature hence are measured in the applied 4-position model of the prototype.

	The other four datasets are man-made features taken from Vienna, the province of Lower Austria, and Austria itself.
	Here we use QGIS 12-position model to measure the conflicts.
	In case of Austria we filtered all places marked as town or city, resulting in a total of 301 features with $ 16.58 $ conflicts per feature.
	For Lower Austria we took all villages, towns, and cities inside the state boundaries of Lower Austria and Vienna. 
	This resulted in a set with  2,260  features with $ 22.6 $ conflicts per feature.
	These settlement features are irregularly distributed according to the physical geography of an alpine country.
	Finally we considered all bus, tram, and subway stops inside the state boundaries of Vienna. %
	These features are more dense in the city center and thin out towards the periphery.
	One set, ``Vienna Train'', consists of 1,001 tram and subway stops with $ 23.27 $ conflicts per feature.
	The last data set we call ``Vienna Bus/Train''.
	It adds also all bus stops inside Vienna to the tram and subway stops. 
	These are  3,939  features and $ 33.66 $ conflicts per feature, hence it is by far the most densely packed set of features.
	Note that for all data sets the number of conflicts includes the conflicts between label candidates of the same feature, which yields a lower bound of 3 or 11 conflicts per feature in the 4- or 12-position model, respectively.

	\begin{table}[tb]
		\caption{Test data sets and their properties.}
		\label{tab:data}
		
		\centering
		\begin{tabular}{lrrrr}
			\toprule
			data set &  features & conflicts & conflicts/feat\\
			\midrule
			Mountain Peaks & 1,278 & 15,416 & 3.01\\
			\midrule
			Austria & 301 & 4,991 & 16.58 \\
			Lower Austria & 2,260 & 47,269 & 22.60 \\
			Vienna Train & 1,001 & 23,294 & 23.27 \\
			Vienna Bus/Train & 3,939 & 132,571 & 33.66 \\
		\end{tabular}
	\end{table}
	
	\section{Semi-Automatic Labeling Prototype}\label{sec:framework}
	We developed a prototype tool that includes four labeling algorithms and provides a proof-of-concept GUI to test user interaction with the system. For the implementation of the backend, especially the algorithms, we used \emph{Java 8} in conjunction with the \href{https://www.playframework.com/}{\emph{Play Framework}}\footnote{\url{https://www.playframework.com/}} 
	(version 2.6) and the \href{https://jgrapht.org/}{JGraphT}\footnote{\url{https://jgrapht.org/}} library 
	(version 1.0.1). For displaying the labels we built a web interface using the Javascript libraries \href{https://leafletjs.com/}{\emph{Leaflet}}\footnote{\url{https://leafletjs.com/}} 
	(version 1.0.3) and \href{https://d3js.org/}{\emph{D3}}\footnote{\url{https://d3js.org/}} 
	(version 4.9.1). 
	\begin{figure}[tb]
		\centering
		\includegraphics[width=\textwidth]{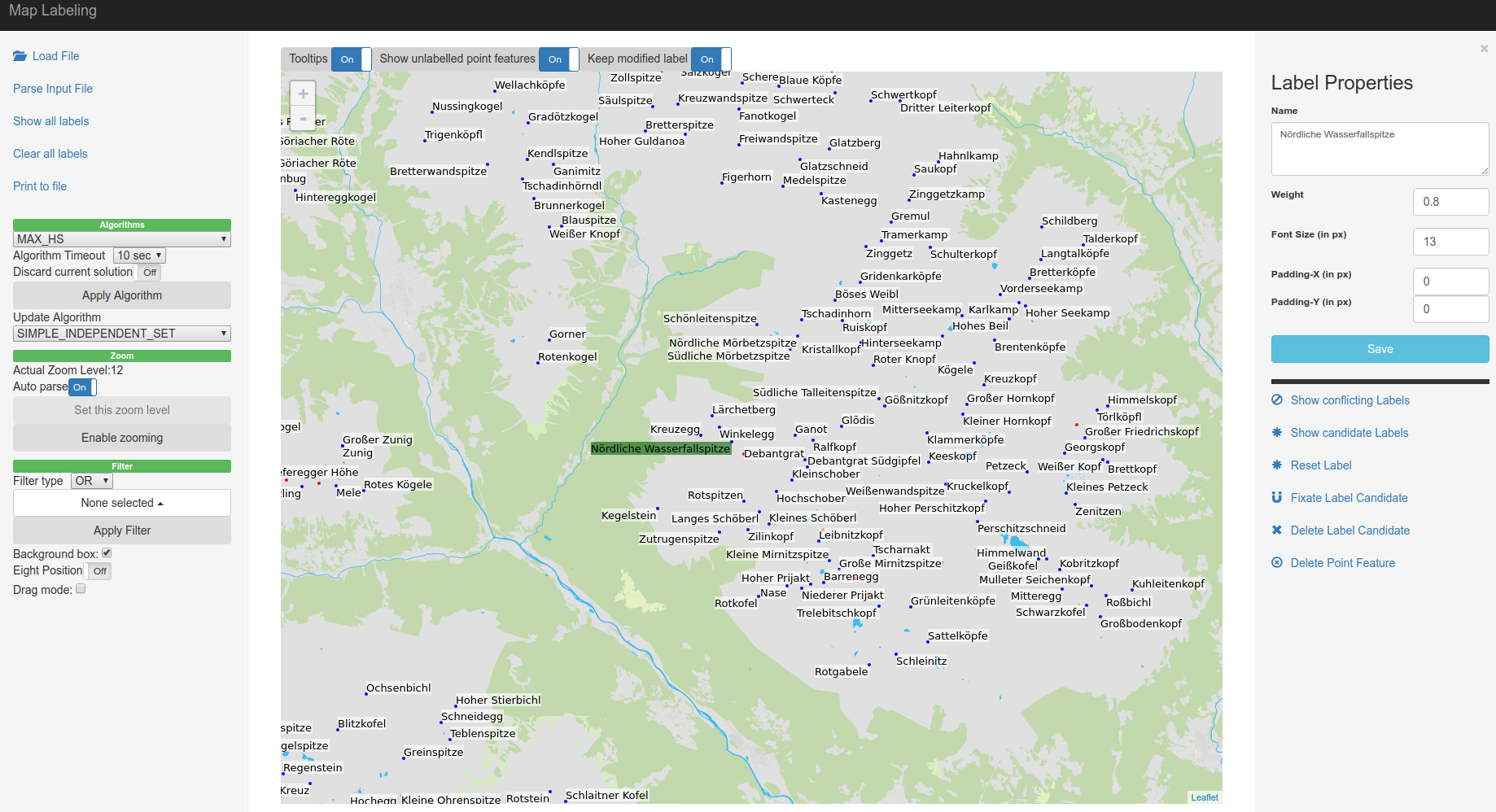}
		\caption{The graphical user interface of the developed prototype.}
		\label{fig:GUI}		
	\end{figure}
	Our application is a one-page design, i.e., the page does not require any reloads while working with it.
	The user interface, shown in Fig.~\ref{fig:GUI}, consists of the large map area in the middle of the window. 
	Here the current labeling is displayed on background map tiles.
	The labels are drawn as white rectangles with black text and are initially attached to the features according to a 4- or 8-position model. 
	All feature points are displayed as filled circles. A blue circle indicates a labeled feature, while a red circle corresponds to an unlabeled one.
	On the left-hand side we find a sidebar, containing most of the input controls, e.g., for loading a file or choosing the algorithms. 
	Above the map area, there are three toggle buttons. 
	Two of them manipulate what is shown on the map, while the rightmost button toggles if the labels we just modified are kept as fixed labels or if they can be deleted from the solution by the next update.
	By clicking on a label, its background color changes to green as seen in Fig.~\ref{fig:GUI} and the label properties area pops up as a sidebar on the right-hand side of the window.
	Here the current values are displayed, e.g., font size, weight, or margin, and the user can modify them accordingly. 
	Shifting a label candidate position by drag-and-drop is also possible.
	Concerning user feedback, the application informs the user via small message boxes on the lower right corner. 
	Longer computations will block the user interface, and a progress bar pops up on the top.

	\subsection{Workflow and User Interaction}
	\label{sec:modification}
	Starting from an unlabeled map, the first step is to import a set of data points to be labeled.
	Then one of the implemented algorithms (see Section~\ref{sec:algorithms}) is selected to compute an initial solution that can subsequently be modified.
	The core of the proposed user-centered labeling process consists of a number of implemented modification tools, that were designed according to the needs of a human cartographer and are summarized in Tab.~\ref{tab:modifications}.
	\begin{table}[tb]
		\small
		\caption{Implemented user modifications.}\label{tab:modifications}
		\centering
		\begin{tabular}{l|l|l}
			
			\toprule
			
			\textbf{text} & \textbf{solution} & \textbf{label}\\
			\midrule
			change font size & delete point features & drag label in the map \\
			change text & delete label candidates & change padding\\
			add line breaks & fixate label candidates & toggle label box visibility\\
			& change cand. weight\\ %
			\bottomrule
			
		\end{tabular}
	\end{table}
	Most of the modifications have a direct effect in the corresponding conflict graph, e.g., deletion or insertion of conflict edges, deletion of vertices, forced selection of vertices, or changes of vertex weights.
	Lastly, the user can select the algorithm for solving the PFLP-Update problem following the modifications; it can be the same algorithm as for computing the initial labeling or a different one, where aspects such as computation speed, stability, and optimality must be taken into account. Our simulation experiments in Section~\ref{sec:statistic} provide some empirical guidance for choosing an update algorithm.

	\subsection{Realistic Sample Workflow}\label{sec:sampleworkflow}
	In this subsection, we describe an example of a realistic map labeling workflow that has been performed by a cartographer using our prototype tool; the process was protocoled and video-captured.
	In the first step, the point features from the Mountain Peaks dataset are added to the map. After loading and parsing the features, the
	user zooms and pans to the area of interest. Once the
	desired zoom level is set, an initial labeling can be produced by any of the four algorithms outlined in Section~\ref{sec:algorithms}. Here we used the exact algorithm \maxhs. 
	After it found a solution, the labeled features are shown with a blue symbol and the corresponding label candidate;
	unlabeled features are indicated by a red symbol.
	
	Usually, a cartographer would now try to group and prioritize features
	according to their attributes first. In this sample scenario, though, we are
	treating all features with the same importance and are only using
	visual cues to manually refine the results of the automatic
	labeling. 
	Reasons for necessary manual adaptations include cases, in which the
	visual connection between symbols and labels is ambiguous or in
	which the label does not correspond well with underlying map
	features (e.g., labels covering lakes).
	While some cases could be avoided by implementing more sophisticated placement rules (e.g., assigning label and feature weights), other cases seem difficult to automate -- either due to their subjective nature or due to the complexity of demands.

	After identifying an improvable label, the cartographic expert would
	click on the label, which activates the view of four (or eight)
	alternative label positions. The preferred label candidate can be
	fixated by marking it and activating ``Fixate Label Candidate'' in the modification sidebar. 
	Alternatively, in ``drag mode'',
	the label can be dragged to the preferred position. 
	Further label
	edits include the option to change the font size and weight as well
	as the label name, e.g., using abbreviations or stacking long label names. %
	Based on the manually selected label, the positions of surrounding labels are re-calculated with the selected update algorithm.

	In our test scenario, about 15 improvable labels (less than 10\% of the initially placed labels) were identified.
	Not all of them had to be changed manually, since updates of neighboring labels and subsequent re-calculations fixed some of the issues. Of course, the instant updates sometimes also resulted in new improvable labels. 
	All in all, the option to select and adjust individual labels while maintaining all automatic labeling functionality was considered highly useful by the expert to speed up the label optimization process in comparison to current cartographic workflows without customized algorithmic support for interactive labeling. %

	\section{Algorithms and Experiments in \qgis}\label{sec:algexpqgis}
	Quantum GIS (QGIS) %
	is one of the most popular open-source geographic information system application in recent years.
	By using the PAL~\cite{ertz2009pal} local search labeling library in its labeling engine, QGIS provides an automated layer labeling.
	The label placement can be customized by choosing labeling algorithms, adjusting styling, etc.
	The newly updated labeling toolbar in QGIS 3 provides more tools for manual label placement, which include changing individual label attributes and moving labels in the map.
	This indicates that the QGIS developers see a clear demand for support of interactive labeling, too.
	Our goal in this section is to investigate the effectiveness and stability of several algorithms (described in Section~\ref{sec:algorithms}) applicable to solve problems PFLP and PFLP-Update, as well as the algorithms from PAL (see Section~\ref{sec:qgis}) that are included in QGIS.
	To this end we integrated our algorithms in the labeling engine of QGIS and performed several simulation experiments. The results are reported in Section~\ref{sec:statistic}.
	
	\subsection{Labeling Algorithms}\label{sec:algorithms}
	We selected four labeling algorithms as representatives of existing labeling approaches, three increasingly sophisticated heuristics and one exact algorithm.
	The simplest algorithm  (called \simple) is an easy-to-implement and fast greedy approach with little to no optimization. 
	The second algorithm (\indset) aims to obtain a good approxima\-tion of the maximum independent set in the conflict graph. 
	Given that a lot of graph libraries provide independent set algorithms it is also very easy to apply by taking an existing implementation. 
	The algorithm \kamis~\cite{kamis2017} computes large independent sets by a combination of several advanced algorithmic techniques such as graph partitioning and kernelization.
	Finally we designed a new exact algorithm (\maxhs) based on a MAXSAT formulation. 
	To solve our MAXSAT formulation we use the solver \href{http://www.maxhs.org/}{MaxHS}, %
	which also gives the name to our algorithm. 
	MaxHS is a freely available solver which ranks highly in the competition held at the annual SAT conferences. %
	
	All algorithms except \kamis support weighted labels, but this does not directly affect our experiments as we initially use unweighted labels.
	Maintaining a previous solution can be done with \simple, \indset and \maxhs. 
	In the case of \simple we just keep the conflict-free subset of the old solution completely and try to extend it.
	For \indset and \maxhs we adjust the weights of the old labels so that they have higher priority to be included in the solution.
	Since \kamis currently does not support weighted instances, prioritizing labels of a previous solution via weights is not possible with \kamis.
	From a theoretical perspective, specialized algorithms that respect a previous solution have been considered recently for the \emph{dynamic independent set} problem~\cite{DBLP:conf/stoc/AssadiOSS18}.  So far, this was not investigated in the light of the PFLP problem.

	\paragraph{\simple}
	Algorithm \simple computes a maximal independent set $D \subseteq V $ in the conflict graph $G=(V,E)$ using a greedy approach. It starts by picking a random vertex $u\in V$ of maximum weight and adds it to $D$. All neighbors of $u$ and $u$ itself are then marked. Next we pick an unmarked vertex $v \in V$ of maximum weight, add it to $D$ and mark $v$ and its neighbors. We repeat this until no unmarked vertex remains. The constructed set $D$ is a maximal independent set of $G$, i.e., it cannot be extended. But there is no guarantee that it is a maximum weight independent set. 
	The selected set of label candidates corresponds to the set $D$.
	Our implementation is also able to take as input an independent set $D' \subseteq V$ and guarantee that for the new solution $D$ we have $D' \subseteq D$. %
	
	\paragraph{\indset}
	Like \simple, the algorithm \indset builds a maximal independent set, but in contrast to \simple it tries to find a good approximation to a maximum weight independent set in $G$. Such approaches are well known in the map labeling literature~\cite{Agarwal1998,Strijk2000}. %
	One way to find a large maximal independent set is to find a small minimal \emph{vertex cover} $D \subset V$ of the conflict graph $G$.
	A vertex cover $D$ has the property that for every edge $(u,v) \in E$ at least one of the two vertices $u,v$ is contained in $D$.
	Consequently, the set $ D' = V\setminus D$ is an independent set, as by definition of $D$ no two vertices in $D'$ can be neighbors in $G$. In our implementation we use the greedy vertex cover heuristic as implemented in the graph library JGraphT. %
	For a vertex $ u $ let $ \deg(u) $ be its degree in the conflict graph, then in each step the algorithm picks the vertex with the smallest ratio of $w(u)/\deg(u)$, adds it to $ D $, and removes $ u $ together with all edges incident to $ u $.

	\paragraph{\kamis} The third algorithm \kamis is based on the maximum independent set solver framework \textit{KaMIS}~\cite{kamis2017}. By combining kernelization, local search, an evolutionary algorithm, graph partitioning and other techniques, this advanced maximum independent solver can very successfully find large independent sets in huge sparse graphs. 
	There are three algorithm components in this framework. Firstly, to create initial solutions, it uses a swap-based local search called \textit{ARW}~\cite{kamis2017}. 
	After a greedy insertion of vertices with small residual degrees in the independent set, the local search applies (1, 2)-swaps. 
	In particular, if two vertices have only one common neighbor in the current solution, the optimization search then inserts these two vertices and removes its neighbor to increase the size of the independent set locally. 
	The second algorithm component is the evolutionary algorithm \textit{EvoMIS}~\cite{lamm2015graph}. 
	It employs a multi-way partitioning on the graph to make the exchange of sub-solutions in graph components possible. 
	After the recombination, newly generated offsprings are locally optimized using swaps. 
	The third component is a kernelization technique~\cite{kamis2017} with both exact and inexact kernels.
	Exact kernelization applies reduction rules to decrease the problem size without affecting solution quality. 
	For example, isolated vertices can always be added to the independent set. 
	Besides exact kernelization, inexact kernelization rules are used to reduce the search space in the optimization phase.
	Intuitively, we choose vertices with very small degree which are in the current candidate independent set and fixate them. 
	Consequently their neighborhood is then deleted, which reduces the size of the remaining instance.
	
	The whole algorithm works as follows. The first phase is to reduce the problem size by exact kernelization. After an initialization phase using \textit{ARW}, the evolutionary procedure \textit{EvoMIS} combines sub-solutions of components and optimizes the newly generated solutions.  Inexact kernelization of the fit solutions make further exact kernelization possible, and the algorithm repeats the whole process from there.
	
	\paragraph{\maxhs}
	Finally we introduce \maxhs, an approach based on satisfiability testing of Boolean formulas. A Boolean formula $\phi$ in conjunctive normal form consists of a logical conjunction of clauses $\phi = c_1 \land \dots \land c_m$.  Each \emph{clause} $c_i$ is a logical disjunction of one or more Boolean variables $x_1, \dots, x_n$ and their negations $\neg x_1, \dots, \neg x_n$; these are called \emph{literals}. Each \emph{variable} can take the value \emph{true} or \emph{false}. For a particular \emph{truth assignment} $\varphi$ of all variables, one can evaluate the truth values of all clauses. The formula $\phi$ is \emph{true} if every clause is \emph{true}. A truth assignment for which $\phi$ evaluates to \emph{true} is called a \emph{satisfying~assignment}.
	
	The satisfiability problem (SAT) asks for the existence of a satisfying assignment, given a Boolean formula $\phi$, and is one of the fundamental NP-complete problems~\cite{Garey79}. If every clause consists of at most two literals, the restricted problem is known as 2-SAT and can be solved in polynomial time~\cite{krom1967decision}. Formann and Wagner~\cite{Formann1991} modeled the labeling problem for a 2-position model as a 2-SAT formula, which allowed them to test in polynomial time whether a conflict-free labeling exists that assigns a label candidate to every point.

	Let $P$ be a set of point features, $\mathcal L$ the label candidates and $G = (V,E)$ the conflict graph of $ P $. 
	We first construct a $2$-SAT formula $\phi$ that guarantees the solution set to be conflict free. 
	Let $\Lambda$ be the set of variables of $\phi$ and $\Gamma$ the set of  clauses. 
	To build our formula, we introduce for each vertex $u \in V$ a Boolean variable $\lambda_u \in \Lambda$, and for each conflict edge $(u,v) \in E$ the clause $ \gamma(u,v) = (\neg \lambda_u \lor \neg \lambda_v) \in \Gamma$. 
	We derive a solution set $\mathcal S\subseteq \mathcal L$ from a truth assignment $\varphi$ of $\phi$, by choosing a label candidate $\ell \in \mathcal L$ to be added to $\mathcal S$ if and only if for the corresponding vertex $u \in V$ the variable $\lambda_u$ is \emph{true} in $\varphi$.
	
	Now $ \mathcal S $ is conflict-free if and only if $ \varphi $ is a satisfying assignment of $ \phi $. This can be seen by remembering that for every conflicting pair of labels we find an edge $ (u,v) \in E $ between the vertices $ u,v \in V $ corresponding to the conflicting labels. For such an edge we introduced the clause $ \gamma(u,v) $ which states that it is not possible to set $ \lambda_u = \lambda_v = \emph{true} $ in any satisfying assignment. While such an assignment leads to conflict-free label sets it does not maximize the number of labels. In particular the assignment $ \varphi $ mapping all variables to \emph{false} is a satisfying one, but the solution set $ \mathcal S $ resulting from $ \varphi $ is empty.
	
	In the related problem MAXSAT we do not ask for a satisfying assignment   of a given formula, but instead for an assignment that maximizes the number of clauses that evaluate to \emph{true}. MAXSAT, as well as MAX-2-SAT, are well known to be NP-complete~\cite{Garey79}.
	To model the PFPL problem as a MAX-2-SAT formula we add a literal for every $\lambda_u\in \Lambda$ as a separate clause $ \gamma(u) = \lambda_u \in \Gamma $. 
	However, if we simply maximize the number of satisfied clauses, we have no guarantee whether some of the clauses %
	$\gamma(u,v)$ evaluate to \emph{false} -- which would imply that two conflicting label candidates can be selected.
	Hence we use a version of the MAXSAT problem, the \emph{partial maximum satisfiability problem} (PMAX-SAT)~\cite{Ansotegui2010}. 
	In PMAX-SAT the set of clauses is partitioned into a set of \emph{hard clauses} $\Gamma_H$ and a set of \emph{soft clauses} $\Gamma_S$. %
	We must find a truth assignment $\varphi$ such that  any clause $\gamma \in \Gamma_H$ evaluates to \emph{true}, while the number of clauses $ \gamma' \in \Gamma_S $ that evaluate to \emph{true} is maximized.
	In our case we define the clauses $\gamma(u,v) \in \Gamma$ %
	as hard clauses and the clauses $\gamma(u) \in \Gamma $ as soft clauses. 
	Now for any solution the clauses expressing a conflict have to be satisfied. 
	We note that a similar formulation has been used to model the maximum clique problem~\cite{flqfx-smwcumsr-14}, which is equivalent to the maximum independent set problem in the complement graph.
	
	As a next step we extend our model to also solve the PFLP-Update problem. 
	In \emph{weighted} PMAX-SAT a soft clause $ \gamma \in \Gamma_S $ can be assigned a weight $ w(\gamma) \in \mathbb{R}^+ $, which can be seen as a penalty for falsifying $\gamma$. 
	The aim is to find an assignment $\varphi$ of $\phi$ that satisfies all the clauses in $\Gamma_H$ and minimizes the sum of penalties of the unsatisfied clauses in $\Gamma_S$.
	Let $ \ell \in \mathcal L $ be a label candidate, $w$ its weight, and $\mathcal S' \subset \mathcal L$ the previous solution. 
	We have to specify the weights for the soft clauses. 
	For every $ \gamma \in \Gamma_S $ that corresponds to a label candidate from $\mathcal S'$ we set $w(\gamma) = w+\varepsilon$ for some parameter $\varepsilon\geq0$, which gives higher priority to selecting a previously displayed label candidate over a previously unused one.
	In our implementation we used $w=1$ and $\varepsilon=1$.
	If we want to strictly prioritize maximum solutions over solutions with fewer labels (but possibly more from $\mathcal S'$) we need to compute a suitable $ \varepsilon $. 
	Let $k = |\mathcal S'|$ and assume all label weights are uniform $w \equiv 1$.
	Then we can choose $0 < \varepsilon < 1 / k$.
	\subsection{QGIS labeling algorithms}\label{sec:qgis}
	For our experiments we further selected three representative algorithms from the labeling engine of QGIS. 
	The \falp algorithm is the simplest greedy algorithm, which is also used to build the initial solution in other optimization algorithms.
	The  algorithm \chain is a local search algorithm with chained neighborhood moves.
	The most advanced algorithm in \qgis is \popmusic (called pop\_tabu\_chain). It combines the general paradigm of \popmusic~\cite{taillard2002popmusic} with \chain and tabu search.
	
	\paragraph{\falp}
	The fast procedure \falp builds an unweighted solution in two steps: First, all label candidates are ordered by  increasing number of conflicts with other label candidates.
	By using the ordered position mode in QGIS, the ordering respects the common preference of candidate positions~\cite{imhof1975} for tie-breaking.
	Then the label candidates are visited in this order. Once a label position is chosen, other candidates of its feature and other label candidates in conflict with this label are removed from the ordering, and all conflict numbers of remaining label positions are updated and re-sorted.
	The implementation in QGIS is incremental. 
	It maintains the conflict number of each candidate and updates the values accordingly.
	
	\paragraph{\chain}
	The local search approach \chain is chaining multiple modifications in order to escape from local minima. 
	After building an initial solution with \falp, improvements of the current labeling are searched by applying a sequence of chained modifications.
	A chain of modification is formed as follows.
	First, a (seed) feature is chosen randomly, and it will be labeled (unlabeled) or modified in the current solution.
	This move may create new overlaps and therefore may lead to a chain of modifications of the current solution.
	The chain search temporarily applies these new changes and the local search process continues.
	The chain search will stop after a specified number of modifications is reached. 
	Once the chain is stopped, the best solution reached along the chain will replace the current one.
	
	\paragraph{\popmusic}
	The \popmusic algorithm is an implementation of the \popmusic framework presented by Taillard and Voss~\cite{taillard2002popmusic}.
	The basic idea applied to the PFPL problem can be summarized as follows: We begin with an initial solution generated by \falp. 
	Every feature is now considered as a sub-part of the instance.
	We also maintain a list $ L $ of features which is initially empty.
	The algorithm iteratively considers features not in $ L $ and executes the \chain algorithm starting from this feature with a bound on the maximum number of labels \chain is allowed to consider in its search. 
	If this procedure leads to an improvement we remove all labels considered by \chain from $ L $.
	In case the run did not further improve the solution we add the feature from which we started \chain to $ L $.
	\popmusic terminates once $ L $ contains all features. 
	Additionally, PAL implements a tabu search paradigm for the \chain procedure to avoid cyclically visiting the same solutions when optimizing from some feature. For further details see~\cite{Alvim2009} and~\cite{ertz2009pal}.

	\subsection{Simulation Experiments}\label{sec:statistic}
	\begin{table}[b]
		\caption{Average running times (in ms) for computing an initial solution (init) or an update (upd), number of initially labeled features (init f), and  number of labeled features on average over all runs (feat). Best values are printed in bold.}
		\label{tab:init}
		\centering
		\resizebox{\textwidth}{!}{
			\begin{tabular}{c|rrrr|rrrr|rrrr|rrrr}
				\toprule
				&\multicolumn{4}{c|}{Austria}&\multicolumn{4}{c|}{Lower Austria}&\multicolumn{4}{c|}{Vienna Train}&\multicolumn{4}{c}{Vienna Train/Bus}\\
				Algorithm & init & upd & init f & feat & init & upd & init f & feat & init & upd & init f & feat & init & upd & init f & feat \\\hline
				\chain & 13 & 13 & \bf 179 & \bf 179 & 140 & 146 & \bf 497 & \bf 534 & 38 & 39 & \bf 282 & \bf 288 & 263 & 266 & \bf 351 & \bf 370 \\
				\hline 
				\falp & 5 & 4 & 176 & 177 & 80 & 79 & 494 & 529 & 19 & 19 & 280 & 285 & 192 & 194 & 349 & 367 \\
				\hline 							
				\simple & \bf 0.9& \bf 0.9 & 152 & 159 & \bf 13 & \bf 14 & 388 & 414 & \bf 3 & \bf 3 & 242 & 248 & \bf 18 & \bf 19 & 277 & 283 \\
				\hline 
				\indset & 14 & 13 & 171 & 175 & 261 & 250 & 457 & 509 & 56 & 57 & 272 & 279 & 701 & 668 & 325 & 347 \\
				\hline 							
				\hline
				\kamis & \bf 123 & \bf 121 & \bf 182 &  \bf 182 &  \bf 581 & \bf 863 & \bf 523 & \bf 560 & \bf 248 & \bf 249 & \bf 289 & \bf 295 & 57746 & \bf 6272 & \bf 373  & \bf 391 \\
				\hline 			
				\maxhs & 181 & 164 & \bf 182 & 179 & 11821 & 4257 & \bf 523 & 537 & 734 & 561 & \bf 289 & 288 & 20937 & 7212 & \bf 373 & 373 \\
				\hline 
				\popmusic & 1599 & 1448 & 181 & 181 & 14684 & 15723 & 504 & 541 & 3915 & 3961 & 282 & 289 & \bf 17820 & 18020 & 355 & 373 \\
				\hline 										
		\end{tabular}}
	\end{table}

	Our experiment focuses on two aspects.
	The first aspect asks how viable our advanced heuristic \kamis and the exact approach \maxhs are compared to the heuristics implemented in \qgis.
	The second aspect asks which combination of algorithms performs best in an interactive scenario as defined by Problem~\ref{prob:PFLP-Update}. 
	All runs of the algorithms produce valid, overlap-free labelings. 
	Our main interest in this experiment lies in the computational performance, the objective value of the solution, and the stability of the updated solutions. We are not claiming that the resulting labelings are competitive with manually labeled maps, as the applied modifications in this simulation study are generated by a random process and not by a cartographic expert.
	Yet the algorithmic performance is expected to be comparable for modifications made purposefully to improve a labeling.
	Nonetheless, two initial examples labelings can be seen in the appendix as Figures~\ref{fig:labelingkamis} and~\ref{fig:labelingpopmusic}. 
	The first one is produced with \kamis, the latter with \popmusic.
	
	In this section, we focus on the data sets for Lower Austria and the denser transport network of Vienna, since they are the largest and most dense label sets, respectively. For the two other data sets our findings are similar.

	All experiments were run on a standard desktop computer with an eight-core Intel i7-860 CPU clocked at $ 2.8 $ GHz and $ 8 $ GB RAM, and running Archlinux kernel version $ 5.1.4 $. We compiled \kamis, as well as our code together with QGIS 3.7, using gcc version 8.3 and cmake 3.14.4. The compile flags for QGIS and \kamis were set to Release. Every test was performed 50 times for each combination of initial and update algorithm.

	\paragraph{Initial Solutions}
	To compare our algorithms with the existing labeling algorithms as found in \qgis,
	we considered four of the datasets presented in Section~\ref{sec:data}.
	Our expectation would be that the greedy heuristics \chain, \falp, \simple, and \indset have clearly faster running times compared to \popmusic, \maxhs, and \kamis,
	while the number of labeled features is expected to be higher for the latter algorithms.
	We present our findings in Table~\ref{tab:init}.
	Our intuitive expectations get largely confirmed for the greedy heuristics.
	\simple is clearly the fastest algorithm, but also leaves a large gap in the solution quality,
	even compared to the other greedy heuristics.
	In terms of running times we have to consider the overhead of building the conflict graph for \indset.
	For the Austria data set this took around $ 11 $ms, for Lower Austria $ 100 $ms, and for the two Vienna data sets it took $ 49 $ms for the sparse and $ 279 $ms for the dense one. 
	If this overhead could be removed, e.g., in an update run we would not need to recompute the full graph each time, \indset would run about as fast as \falp.
	In terms of solution quality we also see that \indset stays behind \chain and \falp.
	Comparing \falp and \chain we see that \falp  runs roughly twice as fast as \chain,
	but the solutions are slightly worse.
	In the end \falp and \chain provide similar results in terms of quality and speed.
	
	For the remaining three algorithms we see that \popmusic can provide pretty good solutions
	in terms of number of labeled features, even compared with the optimal approach \maxhs
	and clearly better ones  compared to the greedy heuristics. 
	For \kamis it turned out that its initial solution was in fact always an optimal solution equivalent to \maxhs.
	Looking at the running times,
	we see that surprisingly \maxhs and \kamis outperform \popmusic by one to two orders of magnitude for all but one data set.
	The data set on which \popmusic surpasses \kamis and \maxhs is also the most dense one, 
	namely the full transport network of Vienna.
	Likely this behavior is due to the fact that especially \kamis uses small cuts in the graphs very well, where
	a cut $ E' \subseteq E $ in a graph $G=(V,E)$ is a set of edges 
	such that $ G $  decomposes into two or more independent parts after removal of $E'$.
	Intuitively the sparser instances also have smaller cuts, while the denser instances lead to more highly connected conflict graphs with large cuts.

	\paragraph{Modifications}
	
	\begin{figure}[tb]
		\centering
		\includegraphics{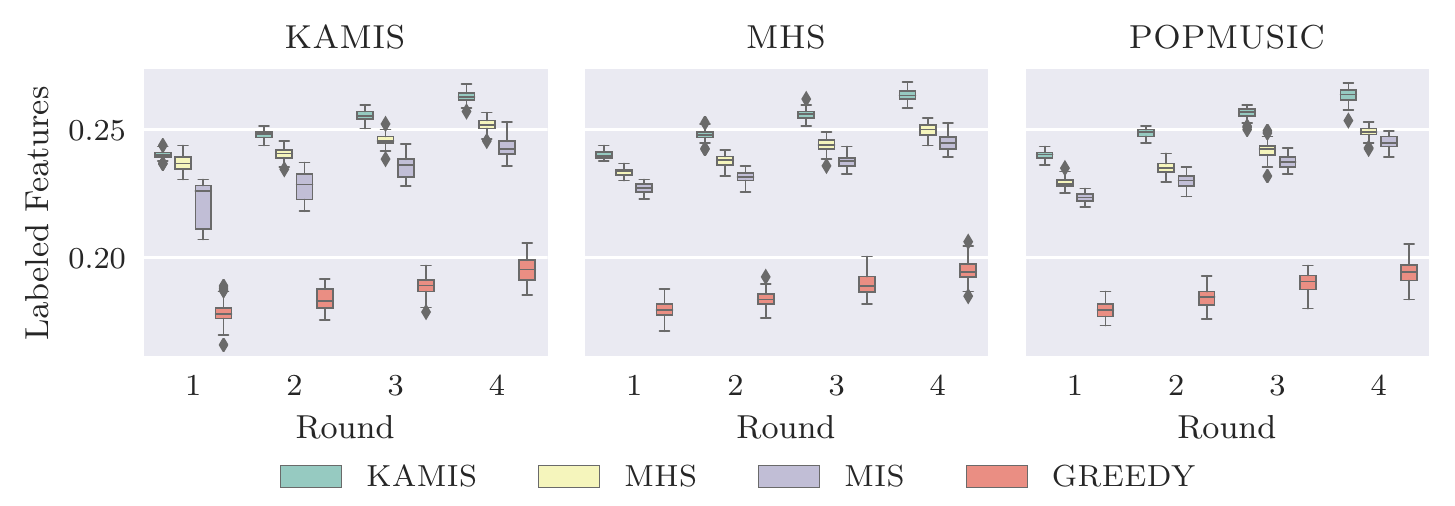}
		\caption{Labeled features in the Lower Austria data set.}
		\label{fig:labeled_LA}
	\end{figure}
	
	\begin{figure}[tb]
		\centering
		\includegraphics{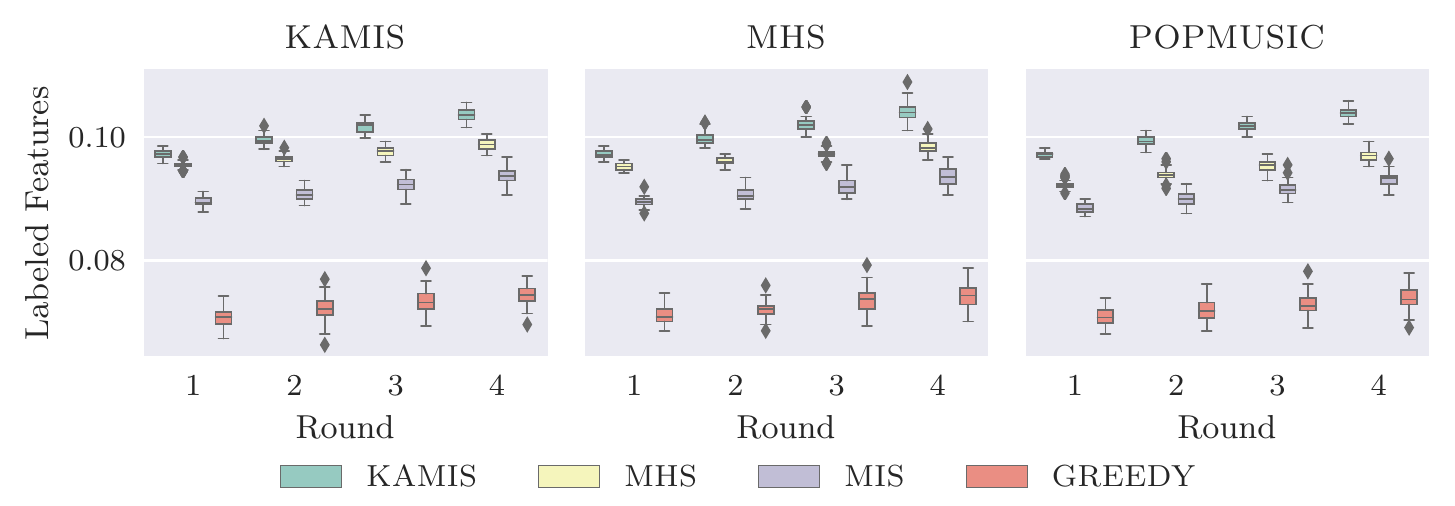}
		\caption{Labeled features in the Lower Austria data set.}
		\label{fig:labeled_VT}
	\end{figure}
	
	For the experiments with modifications we consider five runs of labeling algorithms.
	In the first run we produce an initial labeling.
	As seen in the previous paragraph, \kamis, \maxhs, and \popmusic are the natural candidates for this,
	since they provide the best solution while still running sufficiently fast. 
	To simulate manual modifications of labels in the current solution, we choose after each run labels from the complete set of labels uniformly at random.
	In each round we will change  the font size to $ 20 $ for one percent of the labels, 
	and for another three percent set it to $ 5 $.
	Note that our initial font size is set to $ 10 $.
	Furthermore, we pick one percent of the labels and delete them from further consideration.
	From the perspective of the conflict graph, these modifications cover all relevant changes: insertion and deletion of conflict edges and deletion of vertices.
	It should further be noted that shrinking takes precedence over enlarging, i.e.,
	if we decide to enlarge a label, then shrink it and again enlarge it, it will still remain at a font
	size of~$ 5 $.
	Considering Figures~\ref{fig:labeled_LA} and~\ref{fig:labeled_VT} this explains immediately why on average the number of labeled features increases after each round.
	For computing the updates we are especially interested in \maxhs, \simple, and \indset, 
	as these algorithms consider weights and thus can optimize stability of the previous solution.
	To measure stability of an updated labeling we compute the following ratio.
	Let $ \mathcal S $ be a labeling and $ \mathcal S' $ a labeling of the modified instance,
	then a stable solution keeps $ |\mathcal S \cap \mathcal S'|/|\mathcal S \cup \mathcal S'| $ as close as possible to $ 1.0 $.
	
		\begin{figure}[tb]
			\centering
			\includegraphics{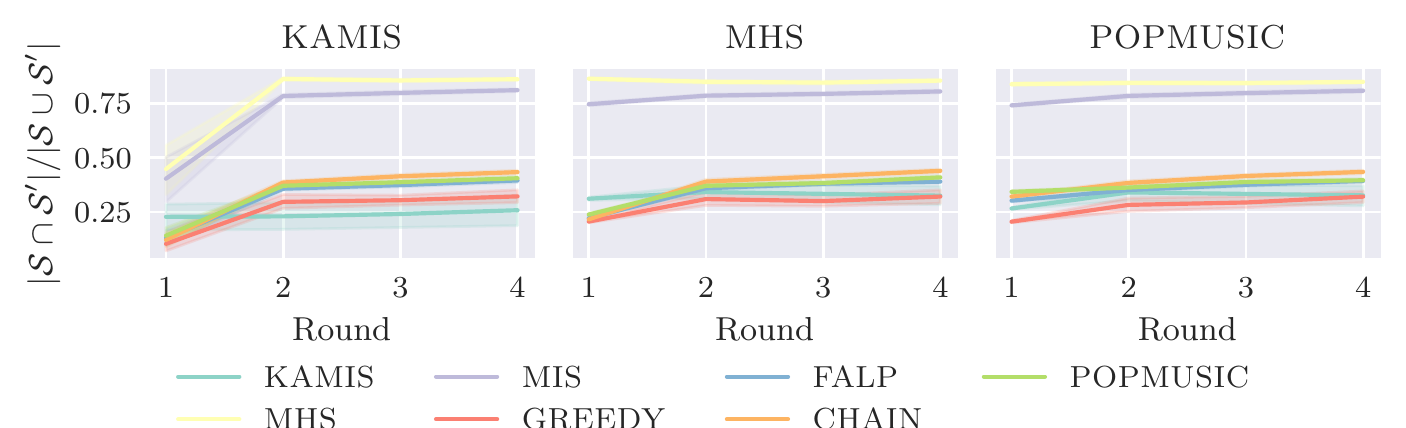}
			\caption{Stability over four rounds of modifications in the Lower Austria data set.}
			\label{fig:stability_LA}
		\end{figure}
	
		\begin{figure}[tb]
			\centering
			\includegraphics{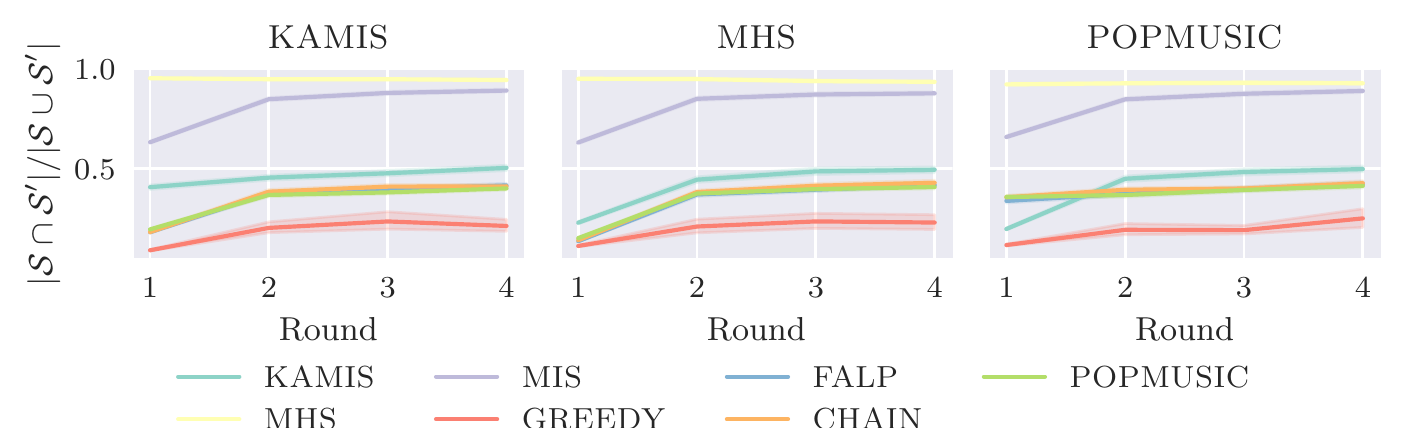}
			\caption{Stability over four rounds of modifications in the Vienna data set of bus, tram, and subway stops.}
			\label{fig:stability_VT}
		\end{figure}
	
	Figures~\ref{fig:stability_LA} and~\ref{fig:stability_VT} show our findings in regard to this measure.
	We used \popmusic, \kamis, and \maxhs as initial algorithms and all algorithms as update algorithms, to explore if by random chance the solutions stay stable as well.
	Clearly we see this is not the case when comparing with \maxhs and \indset,
	as for nearly all possible starting algorithms they manage to keep the ratio at a value of $ 0.8 $ for the transport network of Vienna and $ 0.97 $ for the lower Austria data set.
	In general it seems that \indset is worse at maintaining stability, while \maxhs is computing more stable labeling.	
	Interestingly, for \kamis as a starting algorithm and the dense data set of Vienna \maxhs heavily changes the solution after the first modification.
	This might point to a difference in the solution between \kamis and \maxhs.
	Without further investigation it is hard to determine the exact cause of this change.
	If it turns out that \maxhs and \kamis find very different high quality solutions this could be interesting to cartographers independently of our interactive approach.
	The very naive approach of \simple seems in fact to be too simple to keep the solution stable.
	
	Turning to the running times and number of labeled features, consider Table~\ref{tab:init} again.
	Comparing \maxhs and \indset we see that \maxhs is at a disadvantage in terms of running time,
	but keeps up more labels on average and as just seen produces more stable results.
	On the other hand \indset is very fast and for not too dense input sets seems to be fine in terms of
	stability. 
	
	Since these are average values over all rounds of modification we finally consider boxplots of the percentage of labeled features for \kamis, \maxhs, and \popmusic as starting algorithms and \kamis, \maxhs, \simple, and \indset as update algorithms.
	Our findings are shown in Figures~\ref{fig:labeled_LA} and~\ref{fig:labeled_VT}.
	Remember that \kamis does not try to keep the solution stable, 
	hence in these plots it should rather be seen as a potential optimum for the number of labeled features, 
	if the algorithms were allowed to disregard the old solution.
	As expected we find that \maxhs comes closest to this theoretical optimum.
	\indset is not far behind, while \simple is several percent behind.
	Correlating with the drop in stability we also find that for \kamis as a starting algorithm and \maxhs and \indset as update algorithms we get a large variation in the solution quality on the lower Austria data set.

	\paragraph{Discussion}
	In conclusion we can say that it is more than viable to use exact or near-exact approaches for map labeling when compared with sophisticated heuristics.
	Especially the cut based maximum independent set approach of \kamis is promising in general, as it not only improves the computation times by one to two orders of magnitude over the most sophisticated heuristic \popmusic in QGIS, but also performs consistently better in the optimization goal.
	A combination of \kamis and a heuristic like \popmusic seems like a very useful future approach.
	In such a combined idea one could use cuts to find dense areas which are solved by \popmusic, while the sparser areas are handled by \kamis.
	Also the SAT approach of \maxhs has potential as SAT-solvers continue to get better and the used solver MaxHS surely will be surpassed in the coming years.
	
	In terms of greedy heuristics we saw that \simple and \indset both have their disadvantages.
	\simple is in the end a too trivial approach to realize good solutions,
	while \indset suffers from the overhead of building the conflict graph.
	In a labeling framework though, where the conflict graph would be kept readily available as modifications are made,
	clever independent set heuristics may become a viable approach.
	
	In terms of modifications and interactive labeling we saw that it is possible to use just simple weights to keep the solution stable.
	\maxhs and \indset are both suitable algorithms to handle updates, with \indset clearly being the faster approach.
	A viable combination of the two could be to run \maxhs only every fifth or tenth iteration of modification.
	Also it is likely that future maximum independent set frameworks will support weights, 
	bringing the approach of \kamis also to the table.
	In case these weighted frameworks exhibit a similar running time to \kamis they likely would provide the best of both worlds, fast running times and stable, high quality solutions.

	\section{Conclusion}
	
	We have presented a prototype tool for supporting semi-automatic map labeling workflows together with first experimental results on four possible labeling algorithms and on how to combine them for computing initial labelings and updates after user modifications.
	This is underlined by the QGIS project recently implementing similar ideas into their label placement engine. 
	An immediate consequence from our experiments is that targeted and fast update algorithms that aim for label stability are needed to support interactive modifications.
	In our future investigations we aim to develop algorithms specifically tailored to optimize the  proposed stability criteria.
	As a first step, we plan to investigate fast dynamic weighted independent set heuristics,
	where the weights better model the stability of the labeling.
	Moreover, the labeling algorithms must take into account more advanced and accurate cartographic quality constraints,
	also in combination with the stability criteria.
	Ultimately, this may be fully integrated, e.g., into the QGIS project in order to find its way into practical map production. At that point,  meaningful formal user studies with GIS experts on usability as well as final labeling quality and required interaction efforts are needed to validate whether human-in-the-loop optimization for label placement is meeting its expectations.

	\paragraph{Acknowledgments}{
		This work is
		supported by the Austrian
			Science Fund (FWF) under Grant P31119.
	}

	\bibliographystyle{plainurl}
	\bibliography{paper,martinReferences}
	\newpage
	\appendix
	\section{Supplemental Screenshots}
	\begin{figure}[!b]
		\centering
		\includegraphics[width=\textwidth]{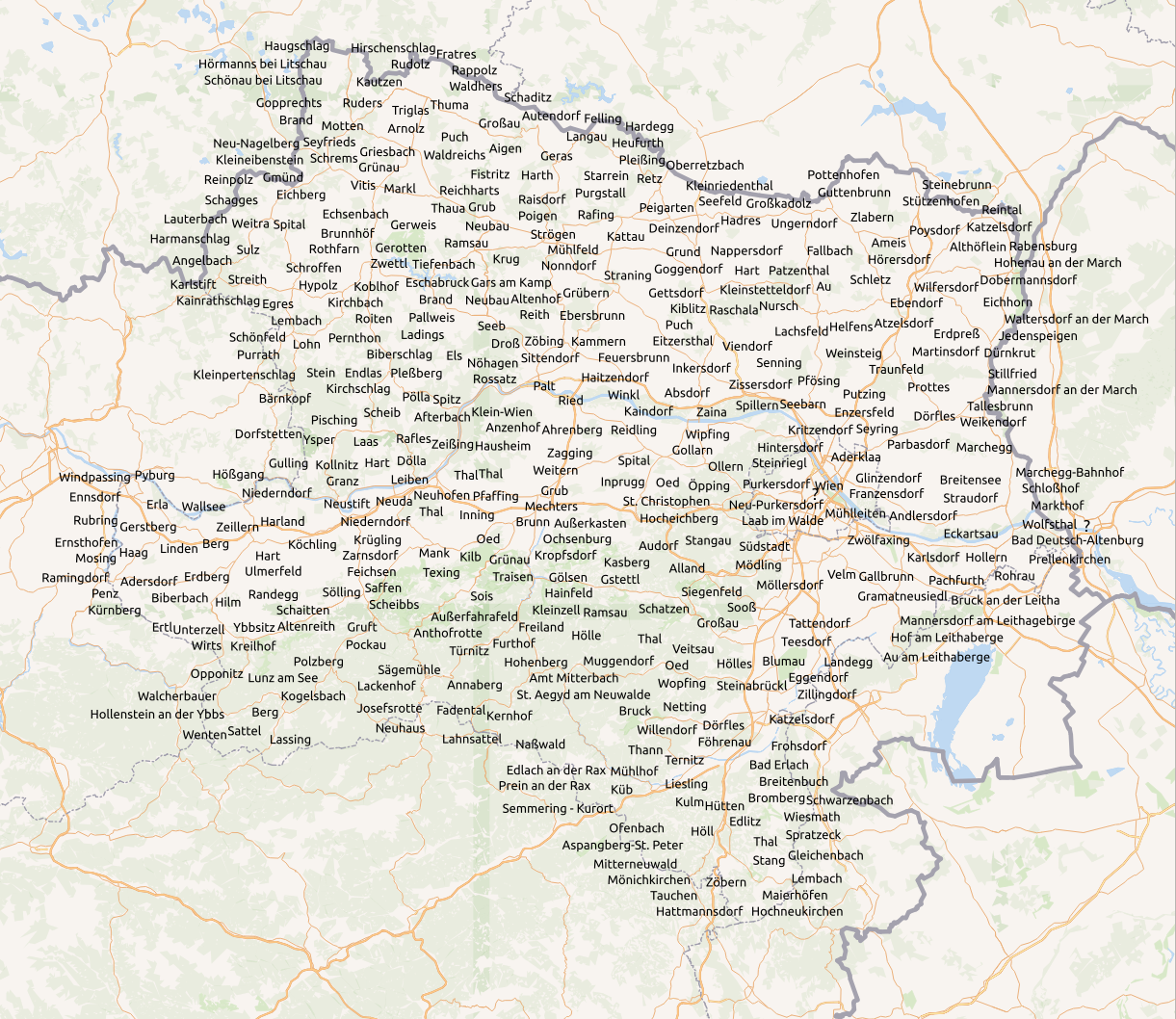}
		\caption{Labeling of the Lower Austria data set computed by \kamis and rendered in QGIS.}\label{fig:labelingkamis}
	\end{figure}
	\begin{figure}[tbp]
		\centering
		\includegraphics[width=\textwidth]{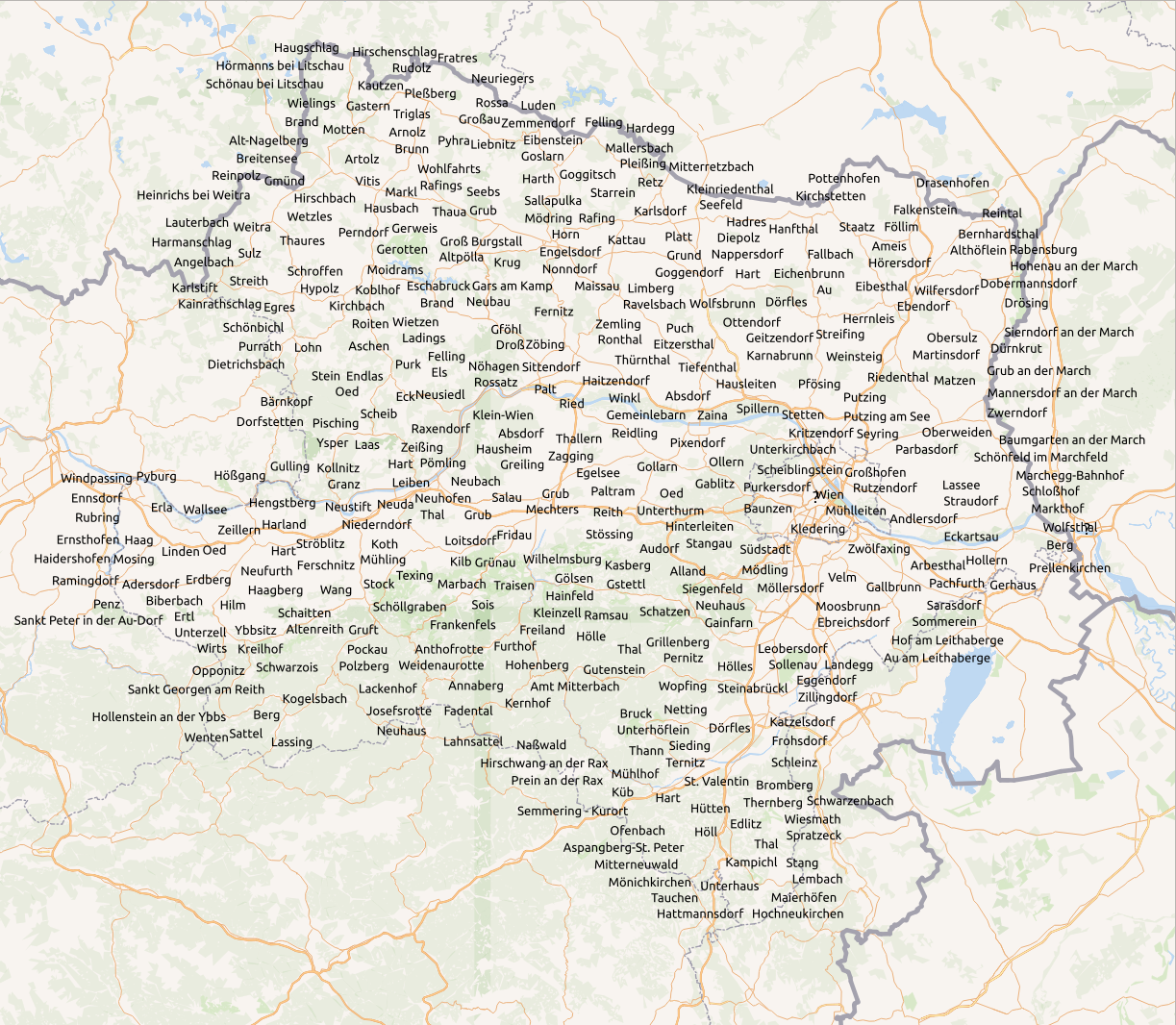}
		\caption{Labeling of the Lower Austria data set computed by \popmusic and rendered in QGIS.}\label{fig:labelingpopmusic}
	\end{figure}

\end{document}